\documentclass[sn-mathphys-num]{sn-jnl}

\usepackage{graphicx}%
\usepackage{multirow}%
\usepackage{amsmath,amssymb,amsfonts}%
\usepackage{amsthm}%
\usepackage{mathrsfs}%
\usepackage[title]{appendix}%
\usepackage{xcolor}%
\usepackage{textcomp}%
\usepackage{manyfoot}%
\usepackage{booktabs}%
\usepackage{algorithm}%
\usepackage{algorithmicx}%
\usepackage{algpseudocode}%
\usepackage{listings}%

\usepackage{array}
\usepackage{longtable}

\graphicspath{{Figures/}}

\theoremstyle{thmstyleone}%
\theoremstyle{thmstyletwo}%

\theoremstyle{thmstylethree}%

\renewcommand{\thetable}{S\arabic{table}}

\begin{document}

\title[Neural models for prediction of spatially patterned phase transitions]{Neural models for prediction of spatially patterned phase transitions: methods and challenges}


\author*[1,2]{Daniel Dylewsky}\email{ddylewsk@uwaterloo.ca}
\author[3]{Sonia Kéfi}
\author[2]{Madhur Anand}
\author[1]{Chris T. Bauch}

\affil[1]{\orgdiv{Department of Applied Mathematics}, \orgname{University of Waterloo}, \orgaddress{\street{200 University Avenue West, Waterloo}, \postcode{N2L 3G1}, \state{Ontario}, \country{Canada}}}
\affil[2]{\orgdiv{School of Environmental Sciences}, \orgname{University of Guelph}, \orgaddress{\street{50 Stone Road East, Guelph}, \postcode{N1G 2W1}, \state{Ontario}, \country{Canada}}}
\affil[3]{\orgname{ISEM, Univ. Montpellier, CNRS, IRD}, \orgaddress{Montpellier},  \country{France}}


\abstract{Dryland vegetation ecosystems are known to be susceptible to critical transitions between alternative stable states when subjected to external forcing. Such transitions are often discussed through the framework of bifurcation theory, but the spatial patterning of vegetation, which is characteristic of drylands, leads to dynamics that are much more complex and diverse than local bifurcations. Recent methodological developments in Early Warning Signal (EWS) detection have shown promise in identifying dynamical signatures of oncoming critical transitions, with particularly strong predictive capabilities being demonstrated by deep neural networks. However, a machine learning model trained on synthetic examples is only useful if it can effectively transfer to a test case of practical interest. These models' capacity to generalize in this manner has been demonstrated for bifurcation transitions, but it is not as well characterized for high-dimensional phase transitions. This paper explores the successes and shortcomings of neural EWS detection for spatially patterned phase transitions, and shows how these models can be used to gain insight into where and how EWS-relevant information is encoded in spatiotemporal dynamics. A few paradigmatic test systems are used to illustrate how the capabilities of such models can be probed in a number of ways, with particular attention to the performances of a number of proposed statistical indicators for EWS and to the supplementary task of distinguishing between abrupt and continuous transitions. Results reveal that model performance often changes dramatically when training and test data sources are interchanged, which offers new insight into the criteria for model generalization.}




\maketitle
\section{Introduction}
Many ecosystems are susceptible to regime shifts between alternative stable states, typically induced when changing environmental conditions cross some critical threshold \cite{Reynolds2007}. These shifts can be abrupt and irreversible, and conventional modeling approaches often struggle to predict if or when such tipping points are approaching. In light of this shortcoming, much research attention has been devoted to the development of Early Warning Signal (EWS) methods which aim to identify dynamical features associated with oncoming regime shifts directly from time series measurements \cite{Dakos2008,Scheffer2009}. Crucially, these EWS indicators are expected to manifest generically across broad classes of critical transitions (albeit with some exceptions and detectability challenges \cite{Boerlijst2013,boettigerEarlyWarningSignals2013,VanderBolt2021}), so they can be utilized even without a model describing the bistable system in question. Empirical studies have applied EWS methodologies in test cases across various domains including ecology, climate, medicine, and social sciences with a fairly high rate of success \cite{dakosTippingPointDetection2024}.

More recently, machine learning methods (in particular, deep neural networks) have been proposed as a powerful tool for EWS detection. These models, which are known to offer state-of-the-art performance in other time series classification tasks \cite{Fawaz2019}, have been shown to produce more accurate and robust predictions of oncoming critical transitions than conventional EWS approaches (albeit at some cost of intepretability) \cite{Bury2021,Deb2022,Dylewsky2023}. One key factor in the success of these ML methodologies is the universality of the signatures they are designed to detect. Training a deep neural network generally requires a large corpus of labeled examples, which can be difficult to obtain from empirical sources. It has been demonstrated, however, that models trained on simple, synthetic (and abundantly available) time series evincing the critical phenomena of interest can be effectively transferred to considerably more complex test cases.

EWS detection has become a topic of interest in dryland ecology, where gradual variation of environmental conditions can induce desertification events in which an ecosystem's vegetated state loses stability and collapses \cite{Scheffer2001,kefiSpatialVegetationPatterns2007}. These regime shifts have a particular theoretical salience because they are strongly mediated by spatial symmetry breaking. The bulk of EWS literature is grounded in bifurcation theory, which offers a minimal explanation of how continuous variation of some parameter in a dynamical system can give rise to qualitative, topological changes in its stability landscape. This framework holds a great deal of explanatory power: many instances of critical transitions even in complex and high-dimensional systems are well approximated by an underlying bifurcation unfolding in some low-dimensional subspace \cite{dylewskyEarlyWarningSignals2024}. However, there are many cases where systems undergo spontaneous regime shifts that are not well characterized by any simple bifurcation form. This broader class of transitions, described by the theory of phase transitions, presents both added challenges and new opportunities with respect to the task of EWS detection \cite{Hagstrom2021}. Dryland vegetation dynamics offer a demonstrative real-world example of how bifurcation theory can fall short: although many spatiotemporal vegetation models do admit mean-field descriptions in which desertification occurs as a local bifurcation, simulation of these models in their full spatial complexity can produce observed transitions very different from those predicted by their mean-field approximations \cite{durrettImportanceBeingDiscrete1994,vonHardenberg2001,rietkerkSelfOrganizationVegetation2002,kefiLocalFacilitationBistability2007}. Practically speaking, the phenomenon of self-organized pattern formation serves to stabilize ecosystems under conditions of environmental stress, delaying or even preventing the complete collapse which would occur in the spatially homogeneous case  \cite{kefiSelforganizationMechanismResilience2024, Rietkerk2021}.

The role of spatial patterning in vegetation regime shifts has complicated the task of anticipating them. The phenomenon of critical slowing down (CSD) that generically accompanies local bifurcations has also been demonstrated to occur in many phase transitions (even, counterintuitively, in those of second order \cite{Hagstrom2021}). Its manifestation in phase transitions can be considerably more complex, however, and statistics such as variance and temporal autocorrelation (which are commonly used as proxies for CSD) are not always reliable as predictors of an approaching transition \cite{Dakos2011}. Fortunately, the comparatively rich diversity of critical phenomena that precede high dimensional phase transitions has enabled the development of a variety of alternative indicators which reflect spatial relationships in the data \cite{Kefi2014}. Many of these are directly analogous to temporal statistics commonly used for EWS detection: just as CSD produces predictable trends in temporal variance, autocorrelation, and power spectrum, so the long-range spatial correlations often exhibited by systems approaching lattice phase transitions lead to similar behavior in the same statistics computed along the spatial dimension(s) \cite{Oborny2005, Guttal2008SpatialVariance, Guttal2008ChangingSkewness, Deblauwe2011, Dakos2009}. Other proposed indicators are more specifically tailored to particular types of spatial organization, such as patchy or periodic patterning \cite{weissmannPredictingCatastrophicShifts2016,manderMorphometricAnalysisVegetation2017}.

Although many such indicators have been devised and evaluated for spatiotemporal phase transitions, the unavoidable challenge of EWS detection for these cases lies in their diversity: whereas certain critical phenomena are guaranteed (with some caveats) to precede local bifurcations, there are no theoretically-established catch-alls that apply across the complete gamut of these irreducibly high-dimensional regime shifts. In practical terms, however, this may not pose a significant obstacle. To establish criteria for oncoming desertification in a dryland ecosystem, for example, one need only consider the subclass of transitions which occur in these particular systems. In this paper we aim to offer some insight into how to operationalize this narrowing of scope in the context of a machine learning approach. Training a neural EWS model requires synthesized training data which is both heterogeneous and specific: ideally, the model is trained on examples of only one type of transition (i.e. that of the test system of interest), and on the most diverse possible array of examples within that type. By ``type" here we mean a class of transitions which exhibit common measurable phenomena as they approach criticality---it is the task of the researcher developing an EWS model for a given application to determine what criteria delineate this class for their particular subject.

The neural model deployed in this work is closely based on similar models which have demonstrated state-of-the-art performance on EWS detection tasks as compared to traditional methods \cite{Bury2021,Deb2022,Dylewsky2023}. Our goal is to offer an example of a modern, adaptable methodology representative of what might be implemented for practical analytic applications. This is not to say that other models of different network architecture (or different algorithms entirely) might not perform comparably well, or even better. However, as model design is not the primary focus of this work, we defer the continued development and comparative evaluation of detection methodologies to other studies.

In the context of dryland vegetation systems, we demonstrate how a machine learning methodology can be used to probe a number of domain-specific EWS questions: How do the various proposed indicators for desertification transitions compare in their predictive capabilities? Which carry the most relevant information, and which offer the earliest warning of oncoming criticality? In addition to detection of these transition events, can we forecast whether the observed shift will be abrupt or continuous? And, perhaps most importantly, how effectively do the learned heuristics generalize to test cases outside the space from which training examples were sampled? We present results obtained using data from a few demonstrative dynamical models exhibiting abrupt and continuous critical transitions. These models do not by any means represent the full range of dynamics across all dryland ecosystems, so our conclusions should not be interpreted as globally definitive answers to these questions; rather, they offer suggestive insights into how different properties of the models affect the presentation of their EWS signatures, and illustrate a useful methodology for the development and refinement of EWS criteria for systems (ecological and otherwise) undergoing spatiotemporal phase transitions.

\section{Methods}
\subsection{Transition Systems}
Three spatiotemporal phase transition models were used to generate the training and test data sets for this study. We selected models with relatively simple governing dynamics capable of generating a critical regime shift resembling dryland desertification (i.e. from a phase in which most lattice sites are occupied to one in which they are mostly empty). Two of these models were developed explicitly with vegetation dynamics in mind, with parameters and interactions that map onto known ecological processes \cite{pichonEstimatingDistancesDesertification2024,weissmannPredictingCatastrophicShifts2016}. The third is the ferromagnetic Ising model on a 2D square lattice, which similarly produces phase transitions mediated by spatial patterning from a local facilitation mechanism. 

Critical transitions in spatially organized vegetation systems can present in sufficiently heterogeneous ways that it would be quite challenging to produce a model library that even approximates their full diversity (see, e.g., \cite{klausmeierRegularIrregularPatterns1999, manorFacilitationCompetitionVegetation2008, vonhardenbergPeriodicScalefreePatterns2010, manderMorphometricAnalysisVegetation2017}). We emphasize that the test models chosen are intended to be representative rather than comprehensive, and while our results offer certain suggestive insights into how spatiotemporal vegetation dynamics can encode EWS information, they should not be understood to reflect properties of all dryland phase transitions. The three models we have selected offer a useful testbed not only because they each generate phase transitions with nontrivial spatial organization from very different defining equations, but also because each has parameter regimes in which the same underlying dynamics are capable of producing continuous or abrupt transitions. This allows for comparisons of EWS detection results across the two types of transitions, as well as tests of classifier models' ability to predict in advance which type is approaching.

\begin{figure}[t]
\centering
\includegraphics[width=\linewidth]{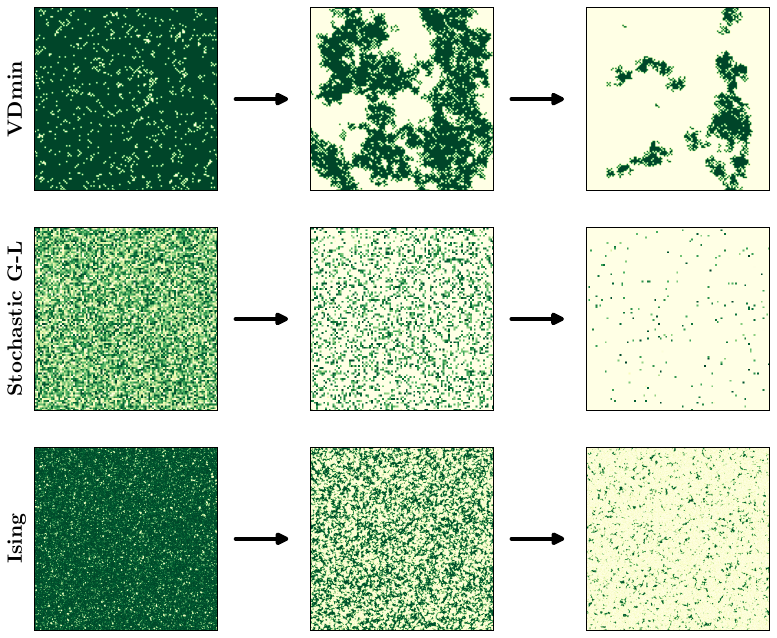}
\caption{Snapshots from simulations of each spatiotemporal model undergoing a critical phase transition.}
\label{fig:trans_model_snapshots}
\end{figure}

The three systems implemented, labeled VDmin \cite{pichonEstimatingDistancesDesertification2024}, Stochastic Ginzburg-Landau \cite{weissmannStochasticDesertification2014}, and 2D Ising, are depicted in Fig. \ref{fig:trans_model_snapshots}. Full descriptions of the models are presented in Table \ref{table:transition_systems} in the supporting materials. Choices of parameters (and parameter ranges) were made to ensure that the resultant time series captured important qualitative features of phase transitions in each model (e.g. that the spatial and temporal scales of their dynamics were reasonable relative to the simulation domain and comparable across models). These selections should not be taken as definitive, however, and in the future it may prove fruitful to extend this analysis to more exotic parameter regimes.

\subsection{EWS Indicators}
\begin{figure*}[t]
\centering
\includegraphics[width=0.9\linewidth]{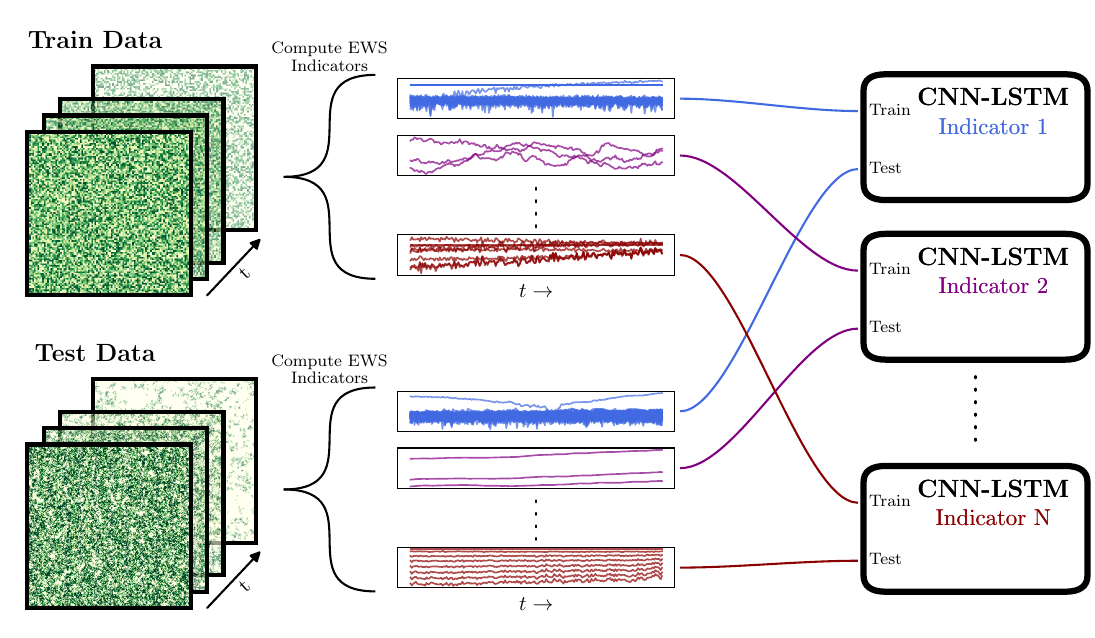}
\caption{Data processing workflow. Time series measurements from spatiotemporal models are dimensionally reduced by computing secondary statistics expected to encode information relevant to critical phenomena. Neural CNN-LSTM models are then trained on the resulting time series data for each of these EWS indicators, and tested on the corresponding statistic computed for the target test case.}
\label{fig:training_workflow}
\end{figure*}

Neural models for EWS detection are not trained directly on the spatiotemporal simulations of the transition systems. Rather, the data is preprocessed by computing various statistical measures which have been suggested to encode EWS-relevant information, following the workflow depicted in Fig. \ref{fig:training_workflow}. One advantage of this approach is dimensionality reduction: larger input dimensions generally mean more trainable model weights, which can lead to overparameterization and overfitting. Although deep neural networks do have a proven aptitude for learning information-preserving reduced-order representations (e.g. in autoencoders), it would be challenging to train for such an objective simultaneously to the optimization of the neural classifier module. Since there already exists a substantial body of research on the development of statistical indicators designed to isolate dynamical signatures of EWS, there is little need to attempt to reinvent these results via machine learning (although this could be a fruitful area for future study). Moreover, training neural classifier models on a variety of such indicators computed on a common set of underlying lattice simulations allows us to compare their relative utility, at least in their application to the chosen test systems. For example, we are able to compare those indicators which are purely temporal, averaged over the observed spatial domain, with those that encode information related to spatial structure. Table \ref{table:ews_indicators} in the supporting materials gives a full description of the 14 indicators implemented in this study, with references to the works in which they were introduced. 

We emphasize that this choice of preprocessing methodology is by no means definitive; models could be trained on any number of other statistical measures, or directly on raw time series data. Our approach is modeled on previous work (\cite{Dylewsky2023}, e.g.) and offers the aforementioned advantages of performance and interpretability, but the goal of this study is not to provide a comparative evaluation of detection models. The results presented do not represent a theoretical performance optimum (no model can guarantee this). Rather, they offer a practical lower bound on the classifiability of the data in question: the measured dynamics encode information sufficient to predict oncoming transitions with \textit{at least} the accuracy attained by our model. The use of secondary statistical measures in data preprocessing produces a dimensionally reduced (compressed) representation of the original data. Information may be lost in this step, but no new information is introduced, so the interpretation of results as a lower bound on performance still holds.

\subsection{Preparation of Training Data}
Training data is generated by numerical simulation of the transition models. Each model can produce abrupt or continuous transitions, depending on choice of parameters, as well as non-transition runs (where the system is never forced near its critical threshold). Examples for each of these categories are simulated in a $2:1:1$ ratio (null : abrupt : continuous). For the sake of maximizing the size and diversity of the training set, each simulation is used as source material for 8 samples, each with randomly selected processing parameters including spatial and temporal crop, spatial downsampling, and measurement noise. For transition runs, temporal cropping is applied such that the time series ends anywhere from 0-400 time steps prior to the identified transition point (and null runs are cropped to a matching distribution). This transition time is computed by numerical estimation of the system-wide inflection point: the spatial average of the dynamics is smoothed with a Gaussian filter, and the transition is deemed to occur when the second derivative of this curve crosses 0. This provides a fairly liberal estimate in the sense that when the time series ends the regime shift is already fully underway, but populating the training set with runs mostly cropped well in advance of this ensures the integrity of the EWS task while also allowing for time-resolved evaluation of the resulting models all the way up to the observed transition.

Thus prepared, these spatiotemporal time series are used to compute each of the EWS indicator statistics. These are then normalized to zero mean and unit standard deviation. Data is separated into training, test, and validation sets (with ratio $0.85:0.10:0.05$), ensuring that all runs derived from the same original simulation are placed in the same class to avoid information contamination. 

\subsection{Implementation of Neural Classifier}
The architecture and hyperparameters of the neural model used in this study are based closely on the designs implemented in \cite{Bury2021} and \cite{Dylewsky2023}. These models concatenate convolutional (CNN) modules with recurrent long short-term memory (LSTM) modules. Input data is passed to two successive 1D convolutional layers containing 20 and 40 filters of width 8 (along the temporal axis), respectively. These feed into a 10\% dropout layer and then a 1D max pooling layer to downsample the temporal dimension by a factor of 2. This output is then passed to an LSTM module containing 20 memory cells. Finally, a dense layer maps results onto the three categorical dimensions associated with the null, abrupt, and continuous classes. Models are trained in TensorFlow using the Adam optimizer and a categorical cross-entropy loss function. After training, a temperature scaling operation is applied such that outputs can be reasonably interpreted as a probability distribution over the classes, as described in \cite{Guo2017}. All models in this study are trained for this three-way classification objective; the sub-tasks of binary classification (transition vs. null) and transition type classification (continuous vs. abrupt) are carried out by algebraic manipulation of the output probabilities (i.e. $P(\text{transition}) = P(\text{abrupt}) + P(\text{continuous})$, or $P(\text{abrupt} |\text{ transition})$ vs. $P(\text{continuous} |\text{ transition})$). 

\section{Results}
\begin{figure}[t]
\centering
\includegraphics[width=0.6\linewidth]{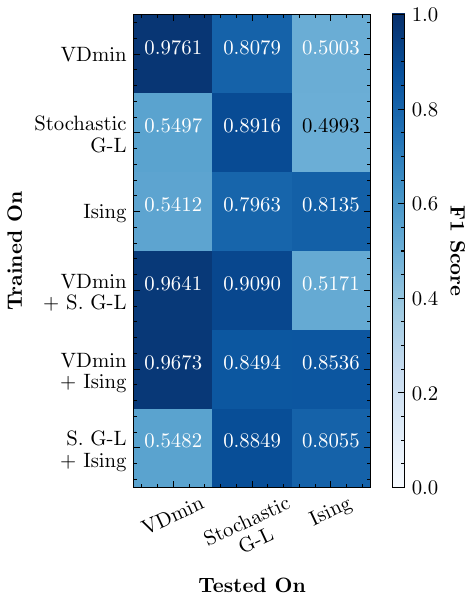}
\caption{$F_1$ accuracies for neural classifiers trained on one or more of the three source systems and tested on withheld data from each model (computed across results for all EWS indicators)}
\label{fig:confusion_matrix_multi_binary}
\end{figure}

Classification accuracies (F1 scores) are reported in Fig. \ref{fig:confusion_matrix_multi_binary} for each CNN-LSTM classifier trained on one or more of the transition models and tested on (withheld) data from each model. Results range from $F_1 \approx 0.5$, which is the expected performance of a completely random binary classifier, to $F_1 \approx 1$, which implies perfect accuracy (precision and recall). Unsurprisingly, accuracies are dependably higher when testing classifiers on data from the same source(s) used in training, and lower when transferring to models not seen during training.

It is immediately apparent that these classifiers are effective predictors of oncoming phase transitions in the system(s) on which they were trained. Performance on models not contained in the training set is less consistent: in some cases results are almost comparable to those for the training system, but in others the neural model fares no better than a random classifier. Interestingly, for the cases where only one model is included in the training set (i.e. the top 3 rows of Fig. \ref{fig:confusion_matrix_multi_binary}), there is no clear symmetry across the diagonal. The accuracy obtained by training on VDMin dynamics and testing on stochastic Ginzburg-Landau runs is significantly higher than that of the inverse case. The success of the former implies that there must exist some shared dynamical signature exhibited by both systems in the lead-up to transition. The poor performance in the latter suggests that either the optimization process taking place during training is failing to identify this signature (perhaps because some other, non-transferrable signature is dominating its classification criteria), or that the signature manifests more weakly in the test data than the training data and thus is not triggering correct classifications.

A neural model which has learned highly system-specific discrimination criteria would likely produce uniformly poor results across all test systems not included in its training, whereas one which has learned more universal criteria would consistently perform well on systems which exhibit those properties strongly and poorly on those where they are weaker. Although our sample of three test systems is not large enough to definitively conclude which of these applies, observed trends strongly suggest that it is the latter: all classifiers not trained on the stochastic Ginzburg-Landau system achieve reasonably high accuracy when transferred to its test set (consistently high F1 scores down the center column of Fig. \ref{fig:confusion_matrix_multi_binary}), but no such uniformity is apparent in each row across test cases outside the training set. This conclusion offers some insight into the observed train/test asymmetry: if the critical phenomena exhibited prior to transition are common to all test systems but manifest more strongly in some than in others, the trained neural models will learn different discrimination thresholds for a positive classification. Supposing that these signatures are comparatively strong in the stochastic G-L dynamics, it stands to reason that the models trained on the other systems would transfer more effectively to these test cases than vice versa.

\subsection{Comparison of Indicators}

\begin{figure}[t]
\centering
\includegraphics[width=0.6\linewidth]{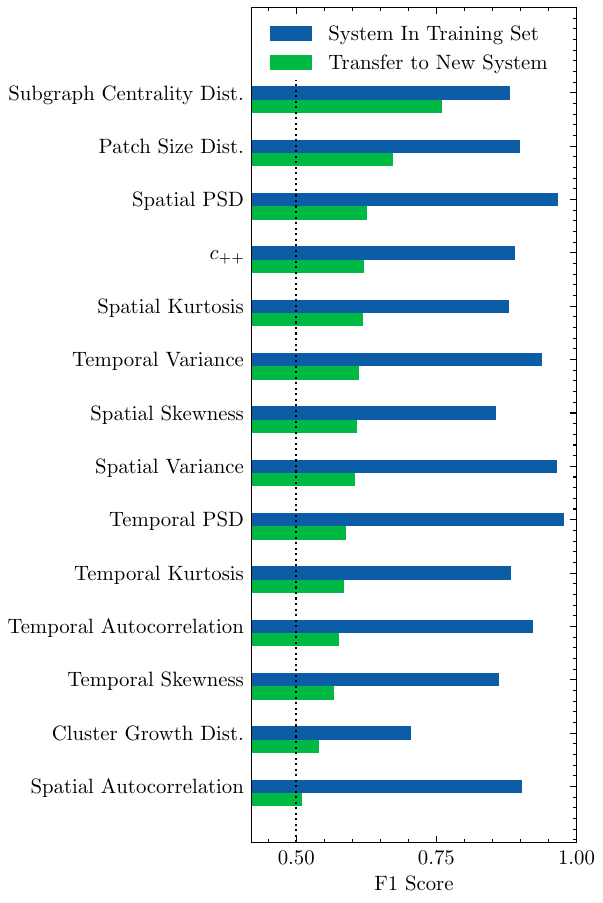}
\caption{$F_1$ accuracy scores for all neural EWS models, separated by the statistical indicators on which they were trained. Performance is reported for tests carried out on simulations of the same system used to generate the training set (green) and for those on simulations of systems not included in training (blue). The black dotted line denotes the expected accuracy of a random classifier, as a baseline.}
\label{fig:Indicator_F1_Scores_binary}
\end{figure}

Classification results for models trained on different EWS indicator statistics are plotted in Fig. \ref{fig:Indicator_F1_Scores_binary}. These scores are computed from results aggregated across all training and test systems. As expected, models consistently achieve higher accuracies when tested on withheld training data than on data from systems not included in the training set. In the transfer case all $F_1$ scores exceed the $0.5$ threshold expected for a random classifier, but some only do so by a small margin. The strongest performances were produced by the subgraph centrality distribution and patch size distribution indicators, which supports the premise that the complex signatures of critical phenomena for spatiotemporal phase transitions can be better detected with more sophisticated, tailored metrics (as opposed to the more commonly used statistical moments and autocorrelation coefficients). 

We wish to emphasize that these results should not be interpreted as a fully general comparative evaluation of the EWS indicators. A few important caveats:
\begin{enumerate}
\item All training and test data was derived from simulations of the three chosen lattice phase transition systems. These systems were selected as representative examples, but they do not by any means cover the full diversity of how such transitions might manifest. 
\item Artificial neural networks are known to be capable of universal function approximation \cite{Hornik1989,Siegelmann1992}, so these models could optimistically be interpreted as functions which make maximal use of the EWS information content of the features on which they were trained. In reality, of course, training takes place on finite data sets over finite duration and there is no guarantee of convergence to this global optimum. For the purpose of measuring the relative effectiveness of these indicators at encoding information relevant to EWS detection, these accuracy scores should therefore be treated as lower bounds.
\item Computation of many of these indicators involves one or more user-defined parameters, and the parameter values which maximize the indicator's utility are often highly dependent on the subject data. For example, the width of the sliding window on which temporal moments are computed should reflect the characteristic time scales of the dynamics in question. For the sake of practical verisimilitude we have defined all such parameters globally across all test systems, so they are not tuned to the dynamical properties of any particular subject. In most cases we do not expect this to significantly degrade performance; however, it may be a determining factor in the poor accuracy score for the cluster growth distribution, which is by far the most highly parameterized indicator tested. We emphasize that this result should not be taken as an indictment of the indicator's utility (which has been demonstrated elsewhere \cite{weissmannPredictingCatastrophicShifts2016}), but at most as a caution against its use in machine learning models intended for generalization.
\end{enumerate}

Even accounting for these limitations, the results in Fig. \ref{fig:Indicator_F1_Scores_binary} offer valuable insight into the manifestation of critical phenomena preceding transitions across all of the test systems. They underscore the extent to which spatial relationships play a key role in signaling oncoming lattice phase transitions (the five top transfer accuracy scores are all spatial statistics), but also the subtlety of how these relationships may be expressed: even though long-distance correlations are commonly observed in lattice systems approaching phase transitions, the spatial autocorrelation statistic performs quite poorly, while the most effective indicators are ones which target more abstract, topologically-oriented spatial properties. This strongly supports the premise that pattern formation is a crucial vector of early warning information for many spatially distributed systems approaching critical transitions.

\subsection{Prediction Horizons}
\begin{figure}[t]
\centering
\includegraphics[width=0.6\linewidth]{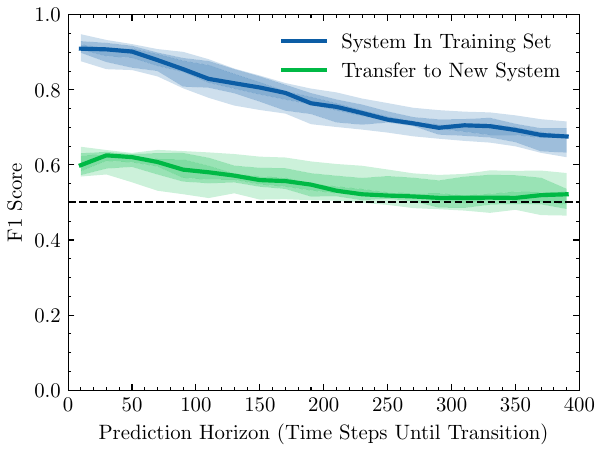}
\caption{Model classification accuracy as a function of time-to-transition. $F_1$ scores are computed across all models tested on withheld training data (blue) and transfer to new data sources (green). The shaded regions indicate the distributions of $F_1$ scores computed separately for each EWS indicator feature, with the solid lines denoting the median values. Time is measured in the arbitrary units of simulation steps.}
\label{fig:F1_vs_ls_binary}
\end{figure}

Model performance is plotted against lead time to the transition in Fig. \ref{fig:F1_vs_ls_binary}. As expected, accuracy increases when the system nears its critical threshold. Interestingly, while accuracy in the transfer case has a maximum effective horizon of about 200 time steps before it drops to the baseline of $F_1 = 0.5$, the tests on withheld training data yield useful predictions out to much greater lead times. When the neural model is trained, we expect it to learn any and all heuristics useful to identifying critical phenomena in the training data. For any given pair of training and test systems, some of these learned dynamical signatures will be common to both (and thus transferable), while others will be specific to the training case. The results in Fig. \ref{fig:F1_vs_ls_binary} illustrate the relative time-resolved forecasting capacities of these two categories (across all pairwise combinations of the phase transition systems used). It is apparent that the inclusion of system-specific classification criteria not only improves overall accuracy, but also offers a considerably longer prediction horizon.

\begin{figure}[t]
\centering
\includegraphics[width=0.6\linewidth]{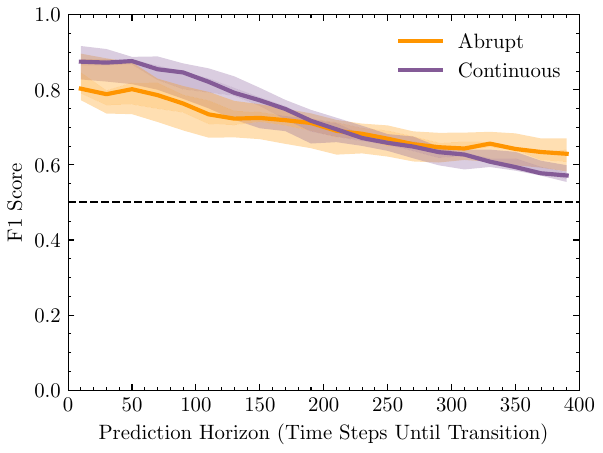}
\caption{EWS classification performance for models tested on withheld training data, separated into abrupt (orange) and continuous (purple) transitions.}
\label{fig:F1_by_trans_type_self}
\end{figure}

Results for tests on withheld training data (i.e. the blue curve in Fig. \ref{fig:F1_vs_ls_binary}) are subdivided to compare the models' ability to predict continuous and abrupt phase transitions across all test systems in Fig. \ref{fig:F1_by_trans_type_self}. Although the disparity between the two is not particularly large, there is a slight advantage in the continuous case for short-term predictions, and likewise in the abrupt case for long-term predictions. This may shed some additional light on previous theoretical work on EWS of phase transitions \cite{Hagstrom2021}, which demonstrated how critical slowing down (CSD) can manifest in each case. Second-order (continuous) transitions exhibit more conventional CSD, analogous to that of a local bifurcation, whereas CSD in first-order (abrupt) transitions is mediated by the system's traversal of metastable intermediary states via spinodal transitions. The results of Fig. \ref{fig:F1_by_trans_type_self} suggest that former mechanism may produce stronger but shorter-lived warning signals (although there may be non-CSD EWS at play here as well).

\subsection{Distinguishing Between Continuous and Abrupt Transitions}

The test systems implemented in this work were chosen for their capacity to produce both continuous and abrupt phase transitions from the same underlying dynamics. This allows for a systematic investigation of the ability of EWS detection models to distinguish between critical phenomena associated with each type of transition. Neural classifier models have shown some aptitude for this task, both for local bifurcations \cite{Bury2021} and lattice phase transitions \cite{Dylewsky2023}. However, while there is ample theoretical support for the existence of early warning signatures which are common across broad classes of critical transitions, it is less clear which (if any) of these signatures can be used to infer the transition type. Given that the methodology of neural EWS detection has typically relied on models trained on synthetic examples being applied to test cases of interest under the assumption that the classification criteria they have learned are transferable, it remains unclear how practical this approach is for the task of differentiating abrupt and continuous transitions. In the context of vegetation systems, it has been suggested that systems exhibiting patchiness induced by strong local facilitation are more likely to undergo abrupt transitions \cite{rietkerkSelfOrganizedPatchinessCatastrophic2004, kefiLocalFacilitationBistability2007}. This property has been successfully exploited to classify oncoming transitions in certain test systems \cite{weissmannPredictingCatastrophicShifts2016}, but the generalizability of this approach is uncertain (particularly in cases where similar spatial pattern formation is not associated with a local facilitation mechanism \cite{martinez-garciaVegetationPatternFormation2013}).

It should be noted that phase transitions observed in real-world vegetation ecosystems cannot always be expected to cleanly reproduce the well-delineated first and second order transitions generated by standard theoretical models. Spatial heterogeneity on all scales leads to dynamics whose medium cannot be perfectly represented as a uniform, isotropic lattice, and resultant phase transitions are not always easy to classify (and could plausibly even be of mixed order and defy categorization altogether \cite{Korbel2025}). Although it is likely that many vegetation systems will, at least on a localized level, undergo transitions which can be well characterized by the binary order taxonomy, the unavoidable messiness of available data makes it difficult to empirically validate a classifier such as ours. Nonetheless, any tool capable of reliably distinguishing abrupt and continuous transitions (even in synthetic data) can offer insight into questions of major practical significance: for a real-world ecosystem approaching some critical threshold, any capacity to anticipate the likelihood of a discontinuous, irreversible equilibrium shift would be extremely valuable. 

\begin{figure}[t]
\centering
\includegraphics[width=0.6\linewidth]{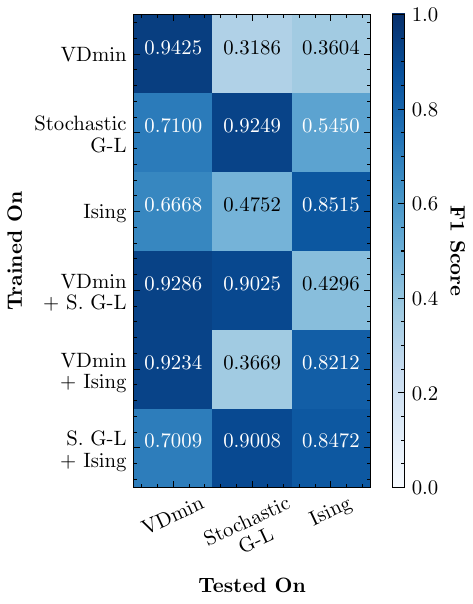}
\caption{Classification accuracies for the task of distinguishing whether an oncoming transition will be continuous or abrupt. Models trained on one or two of the phase transition systems are tested on each one individually.}
\label{fig:confusion_matrix_multi_CA}
\end{figure}

Accuracies plotted in Fig. \ref{fig:confusion_matrix_multi_CA} offer a quantitative evaluation of neural models' performance on this transition type classification across our chosen test systems. These are results for binary classification between continuous and abrupt runs, given that the system is approaching some transition event. Overwhelmingly, we observe high accuracies for models being tested on withheld training data, while models being transferred to new systems are much less consistent. Clearly each of these test systems does present different dynamical signatures preceding abrupt and continuous transitions, but the discrimination criteria as learned by the neural models cannot be expected to generalize across systems. Notably, models trained on two systems have no trouble learning both sets of criteria simultaneously. This suggests that the practical utility of the deep learning methodology for distinguishing abrupt and continuous transitions depends on the user's ability to construct an ensemble of training models with at least one member closely resembling the test case of interest. A sufficiently diverse training ensemble may yield a highly generalizable classifier, but construction of such a data set is no small challenge.

\begin{figure}[t]
\centering
\includegraphics[width=0.6\linewidth]{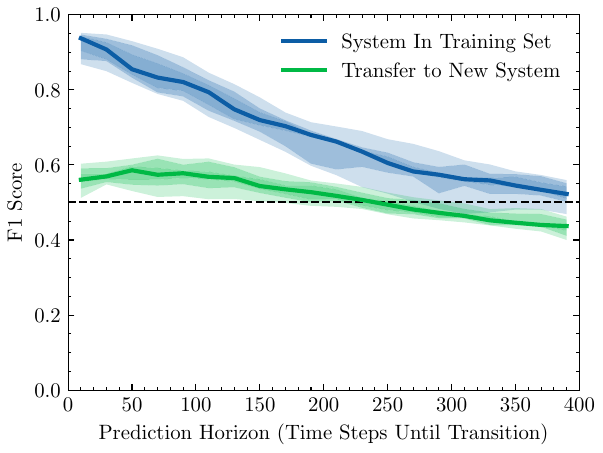}
\caption{Abrupt/continuous classification accuracy as a function of time-to-transition. $F_1$ scores are computed across all models tested on withheld training data (blue) and transfer to new data sources (green). The shaded regions indicate the distributions of $F_1$ scores computed separately for each EWS indicator feature, with the solid lines denoting the median values. Time is measured in the arbitrary units of simulation steps.}
\label{fig:F1_vs_ls_CA}
\end{figure}

\begin{figure}[t]
\centering
\includegraphics[width=0.6\linewidth]{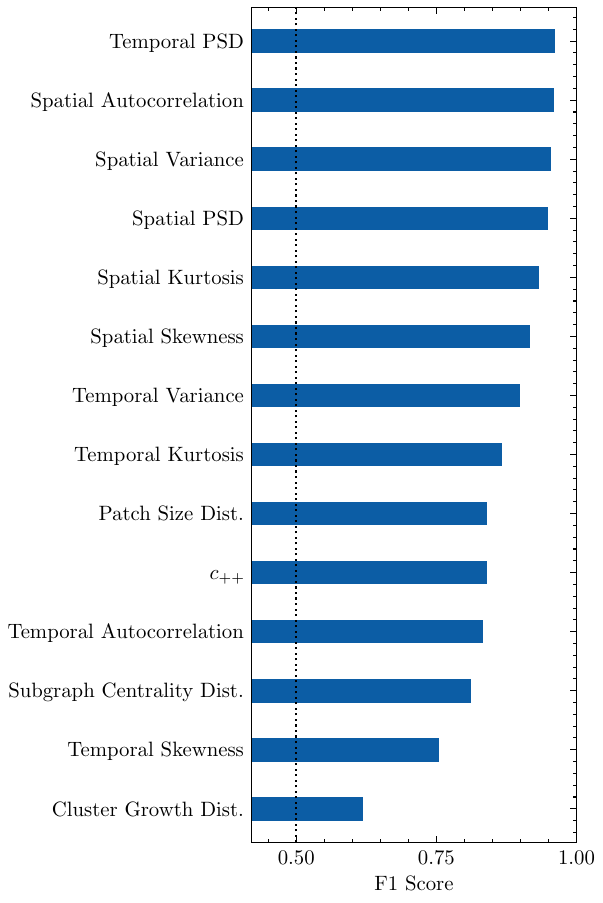}
\caption{$F_1$ accuracy scores for the task of distinguishing abrupt and continuous transitions, separated by the statistical indicators supplied to the classifier. Performance is only reported for tests carried out on simulations of the same system used to generate the training set.}
\label{fig:Indicator_F1_Scores_CA}
\end{figure}

In cases where the neural models do effectively distinguish between the transition types, we can probe their results to learn how information relevant to this task is encoded in the systems' measured dynamics. Fig. \ref{fig:F1_vs_ls_CA} shows the distributions of time-resolved accuracy scores across all EWS indicators. Notably, $F_1$ scores are observed to decrease as prediction horizon lengthens. This illustrates that the models' inferences are relying on properties of critical phenomena which arise near the time of transition (as opposed to dynamical properties which do not depend on proximity to transition). Further insight is offered by comparing performances for each EWS indicator (Fig. \ref{fig:Indicator_F1_Scores_CA}). Examining results for tests on withheld training data (i.e. the blue curve in Fig. \ref{fig:F1_vs_ls_CA}), it is evident that most indicators allow for reasonably high accuracy ($F_1 \approx 0.8$ or greater). Interestingly, there is an opposite trend to that of Fig. \ref{fig:Indicator_F1_Scores_binary}: the conventional EWS indicators (statistical moments, autocorrelations, power spectra) rank more highly, whereas those more tailored to qualitative spatial patterning properties prove less effective. This is even true for the cluster growth distribution, which was explicitly designed to distinguish abrupt from continuous transitions; however, as discussed previously, this may reflect challenges associated with its parameterization rather than the capability of the indicator on the whole.

\section{Model Limitations}
The use of machine learning algorithms, and deep neural networks in particular, inevitably engenders questions of transparency and reproducibility. To an extent, these problems are unavoidable: neural models often offer unrivaled performance capabilities, but this comes at the cost of a near-black box function whose internal workings are very difficult to parse. For the purposes of this study we have attempted to partially circumvent this drawback by adopting an information theoretic lens. Consider some large collection of data samples, sorted into two categories. If the categories were assigned by random coinflip, then any classification algorithm (no matter how sophisticated) could not learn to correctly label a withheld test set of these samples with greater than $50\%$ accuracy. If the algorithm were observed to perform significantly better than this, then, one could conclude that there must be some correlation between the input data and its assigned class. More concretely, the input data must carry at least enough information to support the observed accuracy, regardless of the classification function being used. In this sense, even a black box machine learning model can offer a lower bound on the information content of a data set as it pertains to its training task.

Understanding the classification accuracies presented in this paper strictly as lower bounds on information content offers valuable insight into the test data sets in question, particularly in those cases where models performed notably well. This is the only fully rigorous interpretation of these results in the sense that it is impossible to determine where the neural classifier is underperforming, as compared to a hypothetical, globally optimal model which makes the best possible use of all available information. Less rigorous but perhaps more informative, however, is the interpretation of the reported accuracies as "practical optima." Although there is no guarantee that the neural model we have used is achieving its peak theoretical performance, it does offer a fair approximation of the best accuracy attainable for existing practical methodologies---CNN-LSTM models such as these have demonstrated state-of-the-art performance on a number of EWS detection tasks \cite{Bury2021,Deb2022,Dylewsky2023}. That said, further research would be necessary to assess how robust the reported results are to generalization across different detection algorithms. We suggest that for practical purposes the results presented should be understood as the best available estimates for the corresponding information content (as well as rigorous lower bounds), with the caveat that this interpretation may require updating as methodological research continues to advance.

\section{Discussion}
The results presented demonstrate that deep neural networks have great potential to detect EWS features preceding abrupt and continuous phase transitions in spatial lattice systems. However, models trained on one transition system do not always effectively transfer to test cases from another system (as they have been shown to do with much greater consistency for low dimensional bifurcation dynamics). This is consistent with theoretical expectation: whereas local bifurcations are known to generically produce signatures of critical slowing down as they approach their transition threshold, critical phenomena in the lead-up to high dimensional phase transitions can manifest with much greater diversity. This poses a challenge for machine learning approaches to EWS detection for these systems, which typically rely on classification criteria which can be learned from synthetic training data and transferred to empirical data.

Nonetheless, the measured model performances offer cause for optimism. Although transfer accuracies are inconsistent, a number of train/test model combinations yield strong results. The set of three test systems employed in this work is not large enough to convincingly extrapolate trends of what common features between two transition systems might lead a neural classifier to transfer well between them, but even this limited sampling provides valuable insight into how this problem should be approached. Most notably, the results in Fig. \ref{fig:confusion_matrix_multi_binary} illustrate asymmetry (training on system $A$ and testing on system $B$ does not yield comparable accuracy to training on $B$ and testing on $A$) and recurring directional preference (transfer from other systems onto the stochastic Ginzburg-Landau data consistently produces fairly high accuracies). This implies that the primary obstacle to effective model transfer is not discrepancy between the critical phenomena evinced by the training and test systems, but may instead be the differences in strength and/or timing of critical phenomena common to both systems. If this is the case, it suggests that future machine learning methodologies for EWS of phase transitions should seek to populate their training data sets with systems in which EWS are weak, short-lived, or obfuscated in order to maximize model sensitivity. 

Our study has also offered some insights into the relative utility of proposed EWS indicators as input features for neural classifiers. Although we again emphasize that these results are derived from a limited sample of test systems and should not be taken as universal, the transfer accuracies reported in Fig. \ref{fig:Indicator_F1_Scores_binary} fairly convincingly support the value of indicators designed to measure qualities of spatial patterning associated with lattice phase transitions (particularly subgraph centrality distribution and patch size distribution). However, the differences in accuracy across most tested indicators are relatively small as compared to those between the different train/test model pairs of Fig. \ref{fig:confusion_matrix_multi_binary}, which suggests that the choice of which indicator to use as an input feature is perhaps of secondary importance.

The neural classifiers have also shown some promise for the task of distinguishing between continuous and abrupt transitions, although their performance when transferred to new test data was quite inconsistent. As in the primary classification task (i.e. transition vs. null), there is a pronounced asymmetry between models' accuracy when the training and test systems are swapped. There is again one test system which produces consistently higher accuracies when evaluated by a model not trained on it, but in this case it is the VDmin model rather than the stochastic Ginzburg-Landau. While the decision criteria for the two classification tasks may differ, this phenomenon of directionality appears to be of paramount importance to understanding how generalizable ML EWS detection may be achieved for high dimensional critical transitions.

\backmatter

\bmhead{Supplementary information}

Tables describing the test systems and EWS indicators used are provided in the appendix.

\bmhead{Acknowledgments}
Thanks to the Montpellier Advanced Knowledge Institute on Transitions for funding to MA which initiated and facilitated collaborations with SK.

\section*{Declarations}
\begin{itemize}
\item Funding
The research was supported by NSERC Discovery Grants to MA (5006032-2016) and CTB (5013291-2019), a DARPA Artificial Intelligence Exploration Opportunity Grant to MA and CTB (PA-21-04-02-ACTM-FP-012), and the Canada First Research Excellence Fund (CFREF) Food from Thought Research Program (MA).
\item Conflict of interest/Competing interests (check journal-specific guidelines for which heading to use)
The authors have no competing interests to declare.
\item Ethics approval and consent to participate
Not applicable
\item Consent for publication
Not applicable
\item Data availability 
Not applicable
\item Materials availability
Not applicable
\item Code availability 
Not applicable
\item Author contribution
Study conceived by DD, SK, MA, and CTB. Computational results by DD. Manuscript and figures prepared by DD. All authors reviewed and edited the manuscript.
\end{itemize}

\begin{appendices}

\clearpage
\onecolumn

\section{Appendix}\label{secA1}

\begin{table}
\begin{tabular}{ | >{\raggedright}p{0.15\textwidth} |  p{0.85\textwidth} | } 
  \hline
  \textbf{Model Name} & \textbf{Model Description} \\ 
  \hline
  VDMin & This is a minimal model of vegetation dynamics presented in \cite{pichonEstimatingDistancesDesertification2024} (based on a similar model introduced by \cite{sankaranClusteringCorrelationsInferring2019}). It is a stochastic cellular automaton model which describes a lattice of sites intended to represent vegetation at the scale of individual plants. Desertification transitions are produced by decreasing the plant reproduction rate $p$, which corresponds to an increase in abiotic stress on the system (e.g. environmental aridity). The strength of local facilitation interactions between lattice sites is governed by a second parameter $q$, which determines whether the system undergoes a continuous or abrupt transition. Simulations are carried out based on the protocol outlined in \cite{pichonEstimatingDistancesDesertification2024} on a $128 \times 128$ lattice with periodic boundary conditions. Abrupt and continuous transitions are produced by randomly selecting constant local facilitation coefficients in the strong ($q > 0.7$) and weak ($q < 0.25$) regimes, respectively. \\ 
  \hline
  Stochastic Ginzburg-Landau & This model, introduced in \cite{weissmannStochasticDesertification2014}, is based on the canonical 2D Ginzburg-Landau model for first order phase transitions. Its dynamics are governed by a simple partial differential equation:
\begin{equation}
\frac{\partial b}{\partial t} = D\nabla^2b - \alpha b + \beta b^2 - \gamma b^3
\end{equation}
This can be interpreted as a description of vegetation system where $b$ represents biomass density, $\alpha$ quantifies environmental stress, and $\beta$ is the strength of local facilitation interaction. When $\alpha$ drops below a critical threshold, the system undergoes a first order (abrupt) transition from uniform vegetation ($b>0$) to desert ($b=0$). This simple model is modified by the introduction of environmental noise, the application of which can be tuned to produce first order (weak noise regime) or second order (strong noise regime) transitions. Simulations are carried out using the protocol of \cite{weissmannPredictingCatastrophicShifts2016} on a $100 \times 100$ lattice with periodic boundary conditions. This procedure calls for the application of both multiplicative and demographic noise, but for a more minimal approach we have opted to only use multiplicative noise (which we have found still produces the desired effect). The continuous time integration of the Ginsburg-Landau equation is artificially divided into discrete time steps of length $\Delta t = 0.001$, and the system state is stochastically perturbed once every $\zeta$ steps. Parameters as defined in \cite{weissmannPredictingCatastrophicShifts2016} are randomized on the following uniform intervals: $\beta \in (35,45)$, $\gamma \in (1.4,1.8)$, $\Delta \in (0.3,0.5)$, $D \in (0.15,0.25)$. The integer $\zeta$ is chosen to produce an abrupt transition ($\zeta \in \{32,64\}$) or a continuous transition ($\zeta \in \{1,2\}$).
\\ 
  \hline
  2D Square Ising & This is a canonical phase transition model that has been studied extensively since its introduction by Ising in 1925 \cite{Ising1925}. On a 2D square lattice, it admits both first and second order phase transitions from the same underlying dynamics (achieved by varying external magnetic field and temperature, respectively) \cite{McCoy1973}. Numerical simulations are carried out on a $256\times 256$ lattice with periodic boundary conditions using the Metropolis algorithm, in which the system state is updated asynchronously according to transition probabilities of individual lattice sites. The full system's evolution is measured by snapshots taken once per 10,000 single-site Metropolis iterations. Second order (continuous) transitions are produced by varying temperature $T$ through its critical point with external field $h$ fixed at $0$. First order (abrupt) transitions are produced by varying $h$ with $T$ fixed at a randomized subcritical value. 
\\ 
  \hline
\end{tabular}
\caption{Descriptions of the phase transition systems used to generate training and test data for the neural classifiers.}
\label{table:transition_systems}
\end{table}

\clearpage

\begin{longtable}{ | >{\raggedright}p{0.15\textwidth} |  p{0.7\textwidth} |  p{0.15\textwidth} | } 
  \hline
  \textbf{Indicator} & \textbf{Description} & \textbf{Reference}\\ \hline
  Temporal Variance & Measurement of the second moment of the data about its mean along the temporal axis. Computed on a sliding temporal window (of width 10\% - 40\% of the time series duration) for scalar time series from each lattice site, and then averaged across all spatial coordinates. & Carpenter \& Brock \cite{Carpenter2006}\\  \hline
  Temporal Skewness & Measurement of the third moment of the data about its mean along the temporal axis. Computed on a sliding temporal window (of width 10\% - 40\% of the time series duration) for scalar time series from each lattice site, and then averaged across all spatial coordinates. & Guttal \& Jayaprakash \cite{Guttal2008ChangingSkewness}\\  \hline
  Temporal Kurtosis & Measurement of the fourth moment of the data about its mean along the temporal axis. Computed on a sliding temporal window (of width 10\% - 40\% of the time series duration) for scalar time series from each lattice site, and then averaged across all spatial coordinates. & Xie et. al. \cite{Xie2018} \\ \hline
  Temporal Autocorrelation & Correlation coefficient between successive time steps of the data. Lag-1, -2, and -3 coefficients are computed (corresponding to time lags of 1, 2, and 3 time steps) on a sliding temporal window (of width 10\% - 40\% of the time series duration) for scalar time series from each lattice site, and then averaged across all spatial coordinates. & Held \& Kleinen \cite{Held2004}, Dakos et. al. \cite{Dakos2008}\\  \hline
  Temporal Power Spectrum & Power spectral density of data computed by Fourier decomposition along the temporal axis. Power spectra are computed on a sliding temporal window (of width 10\% - 40\% of the time series duration) for scalar time series from each lattice site, and then averaged across all spatial coordinates. The resulting mean power spectrum is downsampled to its weights across 8 logarithmically spaced bins in frequency space (such that low frequencies are more finely resolved than high). & Kleinen et. al. \cite{kleinenPotentialRoleSpectral2003}\\ \hline
  Spatial Variance &  Measurement of the second moment of the data about its mean along the spatial axes. Computed across the full spatial domain for each time step individually, results of which are then concatenated into a single scalar time series. & Donangelo et. al. \cite{Donangelo2010}, Guttal \& Jayaprakash \cite{Guttal2008SpatialVariance} \\  \hline
  Spatial Skewness & Measurement of the third moment of the data about its mean along the spatial axes. Computed across the full spatial domain for each time step individually, results of which are then concatenated into a single scalar time series. & Guttal \& Jayaprakash \cite{Guttal2008SpatialVariance} \\ \hline
  Spatial Kurtosis & Measurement of the fourth moment of the data about its mean along the spatial axes. Computed across the full spatial domain for each time step individually, results of which are then concatenated into a single scalar time series. Although we are not aware of any specific previous evaluation of spatial kurtosis as an EWS indicator, we have included it for parity with the corresponding temporal indicator. & \\  \hline
  Spatial Autocorrelation & Moran's I spatial correlation coefficient for 1st, 2nd, and 3rd nearest neighbors along the spatial axes. Computed across the full spatial domain for each time step individually, results of which are then concatenated into a single scalar time series. & Dakos et. al. \cite{Dakos2009} \\  \hline
  Spatial Power Spectrum &  Power spectral density of data computed by Fourier decomposition along the spatial axes. The resulting 2D spectral distribution is averaged over the angular coordinate to obtain a radial power spectrum (under the assumption that dynamics are spatially isotropic). This radial spectrum is downsampled to its weights across 8 logarithmically spaced bins in wavenumber space (such that low spatial frequencies are more finely resolved than high). Computed across the full spatial domain for each time step individually, results of which are then concatenated into a single scalar time series. & Deblauwe et. al. \cite{Deblauwe2011} \\  \hline
  Patch Size Distribution & Distribution of sizes of distinct contiguous patches in the spatial domain. To compute this, the input data is thresholded such that each lattice site has a state of 0 (unoccupied) or 1 (occupied) at a given time. The threshold is randomly selected between $\mu-\sigma$ and $\mu+\sigma$, where $\mu$ and $\sigma$ are the global mean and standard deviation of the data, respectively. At each time step, all spatially contiguous patches consisting of one or more occupied lattice sites touching each other are identified. The size distribution of these patches is computed over 8 logarithmically spaced bins spanning sizes from 1 to $2^8$. &  Kéfi et. al. \cite{kefiSpatialVegetationPatterns2007} \\  \hline
  Cluster Growth Distribution & Measurement of the likelihood of a spatial cluster to grow or shrink in time. The same thresholding and patch identification procedure used to compute the Patch Size Distribution is applied, but results for each successive time step are then compared to infer the evolution of the patches over time. A similarity metric derived from \cite{Takaffoli2011, Hartmann2016} is used to determine the continuity of identity of clusters which persist from one time step to the next. Clusters are sorted into 8 size bins, as in the Patch Size Distribution, and probabilities are computed for clusters in each bin to grow or shrink in the next time step. These results are reduced to a scalar (per time step per bin) by taking the mean of $P(\text{grow})$ and $(1-P(\text{shrink}))$. &  Weissmann \& Shnerb \cite{weissmannPredictingCatastrophicShifts2016}\\  \hline
  Subgraph Centrality Distribution & A graph-based measure designed to give quantitative representation to different morphologies observed in spatially patterned vegetation. The same thresholded binary representation used for the previous indicators is treated as a graph in which each occupied lattice site is a vertex and edges connect it to other vertices in its local vicinity. Each vertex is then assigned a subgraph centrality score as outlined in \cite{Mander2013}. The subgraph centralities of all vertices at a given time step form a distribution, which is then sampled into equally spaced bins to form a feature vector of dimension 8. & Mander et. al. \cite{manderMorphometricAnalysisVegetation2017} \\  \hline
  Local Clustering Coefficient ($c_{++}$) & Coefficient to measure the patchiness of a spatial distribution. From the binarized representation of the system, it is computed as the ratio $\rho_{++}/\rho_+^2$, where $\rho_{++}$ is the probability of finding a given pair of neighboring lattice sites both occupied and $\rho_+$ is the overall occupation fraction of the lattice. In this study we use $c_{++} = \rho_{++}/\rho_+^2 - 1$, such that $c_{++} = 0$ when there is no spatial correlation. &  Kéfi et. al. \cite{kefiLocalFacilitationBistability2007}, Iwasa \cite{Iwasa2000}\\  \hline
\caption{Descriptions of the EWS indicators used as input features for the neural models.}
\label{table:ews_indicators}
\end{longtable}

\end{appendices}


\bibliography{sn-bibliography}

\end{document}